\documentclass{kapproc} 

\newcommand{\extraspace}{\addtolength{\abovedisplayskip}{2mm} 
                        \addtolength{\belowdisplayskip}{2mm} 
                        \addtolength{\abovedisplayshortskip}{2mm} 
                        \addtolength{\belowdisplayshortskip}{2mm}} 
\newcommand{\be}{\begin{equation}\extraspace} 
\newcommand{\ee}{\end{equation}} 
\newcommand{\bea}{\begin{eqnarray}\extraspace} 
\newcommand{\eea}{\end{eqnarray}}

\newcommand{\wtPsi}{\widetilde{\Psi}}

\newcommand{\ua}{\uparrow}
\newcommand{\da}{\downarrow}

\newfont{\BBB}{msbm10 scaled\magstephalf}

\newcommand{\BBZ}{\mbox{\BBB Z}}

\DeclareMathAlphabet{\mathbb}{U}{bbold}{m}{n}

\newcommand{\nonu}{\nonumber \\[2mm]} 

\normallatexbib

\begin{document}



\articletitle[Paired and clustered quantum Hall states]
{Paired and clustered \\ quantum Hall states}


\author{K.~Schoutens, E.~Ardonne and F.J.M.~van Lankvelt \\
Institute for Theoretical Physics, Valckenierstraat 65 \\
1018 XE Amsterdam, The Netherlands}



\begin{abstract}
We briefly summarize properties of quantum Hall states
with a pairing or clustering property. Their study employs
a fundamental connection with parafermionic Conformal
Field Theories. We report on closed form expressions for
the many-body wave functions and on multiplicities of
the fundamental quasi-hole excitations.
\savefootnote{Contribution to the proceedings of the NATO Advanced Research
Workshop ``Statistical Field Theories'' Como (Italy), June 18-23 2001}
\end{abstract}



\section{Introduction: new states of matter in the 
         quantum Hall and BEC arenas}

The fractional quantum Hall effect has unveiled states 
of matter that can be characterized as incompressible quantum 
fluids with topological order. Such states are formed in a 
two-dimensional electron gas, at very low temperature and in 
the presence of a strong perpendicular magnetic field.
It has been recognized early on \cite{Halp1,ASW} that 
the excitations over fractional quantum Hall states obey 
fractional braid statistics: a configuration
of $N$ quasi-holes over a fractional quantum Hall ground state forms a 
one-dimensional representation of the braid group $B_N$, where 
the braiding of two quasi-holes is typically represented by 
$e^{i\alpha\pi}$, with $\alpha$ a rational but non-integer number. 
The requirement that particle states have to represent the
braid group rather than the permutation group is special for
two dimensions: the braid group is the fundamental group of
the configuration space of identical particles only in two
dimensions.
On general grounds it is known that, for two-dimensional quantum
systems, higher dimensional representations of the braid group 
$B_N$ are allowed (see \cite{GMS} for an early reference). 
In such a situation, the braiding of particles
is represented by matrices, and since matrices in general do
not commute, this leads to the notion of non-abelian statistics.

It is now believed that the `non-abelian statistics scenario'
is realized in novel types of quantum Hall states, which are 
characterized by a pairing or clustering of electrons under 
quantum Hall conditions. There exists concrete experimental 
\cite{Wi-Pa} and numerical \cite{Mo} evidence that the simplest 
of these states, the `pfaffian state' proposed by Moore and Read 
\cite{MR}, exists in nature, as the state of matter underlying
the quantum Hall effect at filling fraction $\nu={5 \over 2}$. 
At present, this example stands as the lone confirmed `sighting'
of a non-abelian quantum Hall state. 
One expects that paired and clustered quantum Hall 
states can be realized in multi-layer quantum Hall systems with 
sufficiently strong interlayer tunneling, or in situations where 
the spin of the electrons is not fully polarized.

It has recently become clear that quantum Hall states with
clustering correlations are relevant for a very different
class of physical systems, namely Bose Einstein condensates (BEC)
of cold atoms in a rotating trap. In a regime where the rotation
frequency $\Omega$ is of the order of the frequency $\omega$ set 
by the confining potential, and where the healing length
$\xi$ is very large, a rotating BEC becomes formally analogous
to a quantum Hall system of bosonic particles. In 
numerical studies, the quantum liquid that is formed in a 
rotating BEC at vorticity $\nu=N_{\rm boson}/N_{\rm vortex}=k/2$,
has been identified with the Read-Rezayi quantum Hall state
with order-$k$ clustering \cite{CWG}.

In this paper we briefly review the properties of a variety of 
paired and
clustered quantum Hall states, with particular attention for 
states formed by spinful electrons. We present closed form 
expressions for ground state wave functions and discuss the
multiplicities and statistics properties of quasi-hole 
excitations. Throughout our presentation, we stress the role
of parafermionic conformal field theory as a crucial tool
in the theoretical description. 

\section{The fractional quantum Hall effect}

The discovery of the fractional quantum Hall (fqH)
effect \cite{TSG} was truly remarkable and unanticipated.
At fractional filling fraction $\nu=1/3$ a quantum Hall (qH) plateau
was observed. The filling fraction is defined as the
ratio of the number of electrons and the number of available 
states in the lowest Landau level: $\nu= N / N_\phi$,
where $N_\phi$ is the number of flux quanta piercing the 
sample, and $N$ the number of electrons.
Soon after the discovery, Laughlin made a fundamental
break-through by proposing his by now famous wave functions, 
which describe the qH effect at filling fraction
$\nu = \frac{1}{m}$, where $m$ is an odd integer \cite{Laugh} 
\be
\wtPsi_{\rm L}^m(z_1,\ldots,z_N) = \prod_{i<j}(z_i-z_j)^m \ .
\label{eq:lau}
\end{equation}
Here and below we display reduced qH wave functions 
$\wtPsi(z)$, which are related
to the actual wave functions $\Psi(z)$ via $\Psi(z)
= \wtPsi(z) \exp{(-\sum_i \frac{|z_i|^2}{4 l^2}})$ with 
$l=\sqrt{\frac{\hbar c}{e B}}$ the magnetic length. 

Although the qH effect occurs at relatively high 
magnetic fields, it was soon realized that the electron spin
can indeed play an important role. The spin-polarized 
Laughlin states were generalized by Halperin, who proposed a 
set of spin-singlet wave functions \cite{Halp2}
\bea
\lefteqn{ \wtPsi_{\rm H}^{m+1,m+1,m}
(z_1^\ua,\ldots,z_N^\ua;z_1^\da,\ldots,z_N^\da) =}
\label{eq:ss} 
\\ && \quad
\prod_{i<j}(z_i^\ua-z_j^\ua)^{m+1} \prod_{i<j}(z_i^\da-z_j^\da)^{m+1} 
\prod_{i,j}(z_i^\ua-z_j^\da)^{m} \ ,
\nonumber
\eea
where $z_i^\ua$ and $z_j^\da$ are the coordinates of the spin-up
and spin-down electrons, respectively. The state Eq.\
(\ref{eq:ss}) has filling fraction $\nu=2/(2m+1)$. 

\subsection{The qH effect-CFT connection}

\subsubsection{Bulk connection}

Following Moore and Read \cite{MR} one observes that 
it is natural to view (lowest Landau level) qH wave functions 
as conformal blocks of electron-type operators in a suitable 
chiral conformal field theory (CFT) in 2+0 dimensions.
This point of view is related to the fundamental role
of Chern-Simons field theories for qH systems
(compare with \cite{Wi}, where an explicit link between
Chern-Simons theory and CFT is established). 

As an example, the Laughlin 
ground state wave function (\ref{eq:lau}) is obtained as
\be
\wtPsi_{\rm L}^m = \lim_{z_\infty \to \infty} (z_\infty)^{mN^2}
              \langle V_e(z_1) \ldots V_e(z_N) 
              :e^{- i \sqrt{m} N \varphi}(z_\infty): \rangle \ ,
\end{equation}
with $V_e(z)=:\exp (i \sqrt{m} \varphi):$ a chiral vertex
operator in the $c=1$ chiral CFT describing a single
scalar field $\varphi$ compactified on a radius $R^2=m$.


\subsubsection{Edge connection}

While bulk excitations over a qH fluid are gapped,
one expects gapless excitations at the edge of a 
sample. Following Wen \cite{We1}, one observes that the 
edge excitations are described by a chiral Luttinger
Liquid or chiral CFT in 1+1 dimensions. In the example
of the $\nu=\frac{1}{m}$ Laughlin states, one again
has the scalar field theory at $R^2=m$. The neutral
operator $\rho=i\sqrt{m}\partial\varphi$ is associated
with edge density waves, while vertex operators of
type $V^q(z)= :\exp(iq\sqrt{m}\varphi):$ represent charged
edge excitations, the charge being equal to $qe$ with
$-e$ the charge of the electron and $\frac{e}{m}$ 
the charge of the fundamental quasi-holes. 

\subsection{Fractional statistics in the fqH effect}

In the case of an abelian qH state, changing
the magnetic field by one flux quantum $\Phi_0 = \frac{h}{e}$
results in the creation of a quasi-hole (or particle, depending 
on the sign of the change). These quasi-holes can have 
remarkable properties, such as a fractional charge. Also, the
quasi-holes over the Laughlin fqH states are anyons,
i.e.\ they realize fractional braid statistics \cite{Halp1,ASW}.
The fundamental phase for the braiding of two such
excitations is given by $e^{i\frac{\pi}{m}}$.

Closely related to this are the fractional exclusion 
statistics of these same excitations \cite{Hald,ICJ}.
Focusing on edge excitations, one can show that the 
gapless, charged edge excitations of an
abelian qH state satisfy a form of exclusion statistics
closely related to that of Haldane \cite{Hald}. A particularly
natural choice of basis for the edge excitations
employs edge electrons and quasi-holes as the 
fundamental quanta \cite{vES}. In this basis, the exclusion 
statistics parameter matrix is diagonal with self-exclusion
parameters equal to $m$ (for the edge electrons) and 
$\frac{1}{m}$ (for the edge quasi-holes).

For general abelian qH states, one may argue \cite{ABGS}
that the statistics matrix $G$ of edge excitations (in
a specific basis) is of the form
\be
{\bf G} = {\bf K}_e \oplus {\bf K}_e^{-1}  ,
\label{eq:GKK}
\end{equation}
where ${\bf K}_e$ is the so-called $K$-matrix that
characterizes the topological order of the qH
state (see for instance \cite{We2}).

\section{Paired and clustered qH states}

Prompted by the observation of a qH effect at filling fraction 
$\nu=\frac{5}{2}$ \cite{Wi-Pa} 
a number of novel qH states have been proposed. 
Among these is the Moore-Read (MR) state or pfaffian state, which is 
characterized by a $p$-wave pairing of the 
electrons \cite{MR,GWW}. A generalization, where the pairing 
is replaced by a clustering of order $k$ was proposed by Read 
and Rezayi (RR) \cite{RR2}. In \cite{AS1} two of the present authors made 
a further generalization to a class of spin-singlet qH 
states also characterized by a clustering into $k$-plets of 
electrons.

\subsection{Spin-polarized states}
The wave function of the (spin polarized) MR 
state is given by
\be
\wtPsi_{\rm MR}(z_i) = {\rm Pf} (\frac{1}{z_i-z_j})
\prod_{i<j} (z_i-z_j)^{M+1} \ ,
\end{equation}
where ${\rm Pf} (M_{i,j}) = \frac{1}{2^{N/2}(N/2)!} \sum_{\sigma}
\mbox{sgn} (\sigma) \prod_{r=1}^{N/2} M_{\sigma(2r-1),\sigma(2r)}$
is the pfaffian of an antisymmetric matrix.
For the wave function to be antisymmetric (we are describing 
electrons) $M$ needs to be an odd integer, which implies an 
even-denominator filling fraction $\nu = \frac{1}{M+1}$.

The MR wave function can be viewed as a correlator in a 
$c=3/2$ CFT, consisting of a free scalar field and 
a Majorana fermion. The electron operator becomes 
$\psi(z):\exp(i \sqrt{M+1}\, \varphi_c):(z)$.
The correlator of $N$ electron operators (and a 
suitable background charge) splits into a product of vertex 
operators, giving the Laughlin part of the wave function, and
a product of fermion fields, which gives the pfaffian factor. 

Upon generalizing the Majorana fermion to the $\BBZ_k$ 
parafermions \cite{ZF} associated to the coset 
$\frac{\widehat{{\rm su}}(2)_k}{\widehat{{\rm u}}(1)}$, 
one obtains the clustered states of \cite{RR2}. Their wave 
functions are constructed in the same way as the MR 
wave function, with explicit parafermion factors brought
in by the electron operator. The result is a state in which 
the electrons form clusters of order $k$ rather than pairs.  
The filling fraction takes the form $\nu=\frac{k}{kM+2}$, 
with $M$ an odd integer. Note that for $k=1$ the Laughlin
states (with $m=M+2$) are recovered, while $k=2$ gives
the MR states. 

In ref.~\cite{CGT}, Cappelli {\it et al.}\ presented a particularly 
simple construction for $k$-clustered qH states, and they 
showed that the resulting wave functions are equivalent to the 
ones originally obtained by Read and Rezayi \cite{RR2}. 
In this construction, one starts with an abelian state of $k$ 
types of discernible electrons, with the same total
filling fraction as the clustered states. For the RR states, this
abelian state takes the form \cite{CGT}
\be \label{abrr}
\wtPsi_{\rm ab} = \prod_{a=1}^k \prod_{i<j} (z^a_i-z^a_j)^2 \ ,
\ee
where $a$ labels the different types of electrons.
In the construction, we will assume that $M=0$; the wave function
for $M\neq 0$ are obtained by multiplying the final result with the
Laughlin factor $\prod_{i<j} (z_i-z_j)^M$.
Note that the wave function (\ref{abrr}) 
can be thought of as $k$ copies of the clustered state with
parameters $k=1,M=0$ (which is a bosonic Laughlin state at filling 
fraction $\nu=\frac{1}{2}$). The clustered states
are obtained by effectively making the electrons indiscernible, which
is achieved by symmetrizing the expression (\ref{abrr}) over all
the electrons. The wave function for the RR states 
can now be written in the following form
\be \label{rrcapf}
\wtPsi_{\rm RR} = 
\mathcal{S} \left[ \prod_{a=1}^k \prod_{i<j} (z^a_i-z^a_j)^2 \right] 
\prod_{i<j} (z_i -z_j)^M \ ,
\ee
where it is assumed that after the symmetrization procedure all the 
electrons are labeled by a single index.

The clustering property of the wave function (\ref{rrcapf}) for $M=0$
is inherited from the abelian state (\ref{abrr}).
The clustering property
of order $k$ means that the wave function does not vanish if up
to $k$ particles are brought to the same position; however, if 
$k+1$ particles are at the same position, the wave function does vanish.
To see this, note that $k$ electrons all of different type are brought
to the same location, the wave function (\ref{abrr}) does not vanish.
The symmetrization makes sure that this property holds for any $k$
electrons of the clustered state (\ref{rrcapf}) (with $M=0$).

\subsection{Spin-singlet states}

The clustered states discussed in the previous section do not include 
a spin degree of freedom. As was shown in \cite{AS1} and \cite{ALLS}, 
the clustering property can be extended to spin-singlet states of 
spinful electrons, leading to what have been called `non-abelian 
spin singlet' (NASS) states. As stated before, the Halperin states 
are spin-singlet
analogues of the Laughlin states. In the same way, the NASS states of
\cite{AS1} are spin-singlet analogues of the RR
states. To construct their wave functions, two boson fields 
are needed (for charge and spin) in addition to the `higher rank'
parafermions associated to 
$\frac{\widehat{{\rm su}}(3)_k}{[\widehat{{\rm u}}(1)]^2}$
\cite{Ge}. The NASS states have filling 
fraction $\nu=\frac{2k}{2kM+3}$ with $M$ an odd integer.
For $k=1$ the Halperin states (\ref{eq:ss}), with $m=M+1$, 
are recovered. 
Explicit expressions for the NASS wave functions of \cite{AS1} can 
be found in \cite{ARRS}. The form of the wave functions presented
there is similar to the explicit RR wave functions in
\cite{RR2}. Alternatively, a formulation similar to the construction
for the RR states presented above is also possible. 
We again start with $k$ copies of the $k=1,M=0$ states, which are,
in this case, the Halperin states with labels $(2,2,1)$. Again, this
is a bosonic state. To obtain the fermionic states, one has to multiply
in the end with an overall Laughlin factor $\prod_{i<j}(z_i-z_j)^M$ with
$M$ odd. The abelian state from which the clustered spin-singlet states
can be obtained takes the form
\be \label{abas}
\wtPsi_{\rm ab} = \prod_{a=1}^k
\left(
\prod_{i<j} (z^{\ua,a}_i-z^{\ua,a}_j)^2 
\prod_{i<j} (z^{\da,a}_i-z^{\da,a}_j)^2
\prod_{i,j} (z^{\ua,a}_i-z^{\da,a}_j) \right) \ .
\ee
Again, we have $k$ types of electrons, but now also the spin is taken
into account. It is not so hard to find the filling fraction of this
abelian state, which is given by $\nu = \frac{2k}{3}$, which is indeed
the filling fraction of the NASS states with $M=0$. 
To obtain the clustered state, one has to symmetrize the 
wave function (\ref{abas}). However, because we are dealing with
spin-singlet states, the symmetrization procedure is more involved than
for the RR states. The conditions which are to be satisfied by the
spatial part of the wave function in order be a spin-singlet are
the Fock cyclic condition, and the conditions that the wave wave function 
be symmetric in the spin-up and spin-down coordinates separately. 
In general, this can be achieved by the following (generalized) 
symmetrization procedure (see \cite{ARRS} for more details). First
antisymmetrize in $z_1^\ua,z_1^\da$, followed by antisymmetrization
in $z_2^\ua,z_2^\da$, etc. After the antisymmetrization of 
$z_{N/2}^\ua,z_{N/2}^\da$, the wave function has to be symmetrized over
all the spin-up coordinates, and finally, over the spin-down coordinates.
This gives rise to clustered spin-singlet wave functions. Though we do not
have a proof, we believe that via this procedure, we obtain wave functions
which are equivalent to the ones described in \cite{ARRS}. For the case
at hand, the antisymmetrization steps as described above do not seem to 
be necessary (as was the case for the explicit formulation in \cite{ARRS}).
Thus we arrive at the following form for the wave functions of the
clustered spin-singlet states
\bea \label{ascapf}
&&\wtPsi_{\rm NASS} = \nonumber \\ 
&&\mathcal{S}_{z^\ua,z^\da}
\left[ \prod_{a=1}^k \left( \prod_{i<j} (z^{\ua,a}_i-z^{\ua,a}_j)^2 
\prod_{i<j} (z^{\da,a}_i-z^{\da,a}_j)^2
\prod_{i,j} (z^{\ua,a}_i-z^{\da,a}_j) \right)
\right] \nonumber \\
&&\times \prod_{i<j} (x_i -x_j)^M \ ,
\eea
where $x_i$ stands for either $z_i^\ua$ or $z_i^\da$.
$\mathcal{S}_{z^\ua,z^\da}$ denotes symmetrization over the spin-up and 
spin-down particles. The clustering property of these spin-singlet wave
functions can again easily be read off from this formulation. For more
details, see \cite{ARRS}.

Recently, another type of clustered spin-singlet qH states
was proposed \cite{ALLS}. These states are based on parafermions
corresponding to another rank 2 affine Lie algebra, namely the
$\frac{\widehat{{\rm so}}(5)_k}{[\widehat{{\rm u}}(1)]^2}$ parafermions.
Also, two chiral boson fields for the charge and spin degrees of freedom
are present. At level $k=1$, the only parafermion field present in this
theory is the Majorana fermion, which also appeared in the theory for
the MR state. It is thus to be expected that the state corresponding to
the level $k=1$ affine Lie algebra $\widehat{{\rm so}}(5)$ is related to
the MR state. This is indeed the case; the wave function also contains
a pfaffian factor \cite{ALLS}
\be \label{scsepwf}
\wtPsi_{\rm SCsep}^{k=1} =  {\rm Pf} (\frac{1}{x_i-x_j})
\wtPsi_{\rm H}^{M+1,M+1,M} (z^\ua_i,z_j^\da) \ ,
\ee
where $\wtPsi_{\rm H}^{M+1,M+1,M}$ is the Halperin wave function
(\ref{eq:ss}). For $M$ odd, this describes a qH system
at filling $\nu = \frac{2}{2M+1}$. 
This wave function is different from the previous
ones in a few respects. First of all, the (fundamental) excitations
over this state show a separation of the spin and charge degrees of
freedom (SCsep). This is a consequence of the structure of the underlying
Lie algebra symmetry, as was pointed out in \cite{ALLS}. In effect, 
there are spinon like excitations, with spin-$\frac{1}{2}$ and no
charge, and holons which carry charge $\frac{1}{2M+1}$ without spin. 

The other feature which is different in comparison to the clustered states
(\ref{rrcapf}) and (\ref{ascapf}) is the clustering property.
The clustering of the `spin charge separated state' (\ref{scsepwf}) 
(with $M=0$), is in  fact a clustering of the spin-up and spin-down particles
separately. Note that the wave function for $M=0$ has poles when spin-up 
particles are at the same location as spin-down particles. 
Nevertheless, we will discuss the clustering for the case $M=0$.
In the case $k=1$, up to two particles of the same spin can be brought 
to the same location while the wave function is still non-zero. If we 
first bring $z^\ua_1,z^\ua_2 \to z^\ua$ and $z^\da_1,z^\da_2 \to z^\da$, 
and then send $z^\ua \to z^\da$ the wave function remains non-zero. 
The clustering is thus different from the clustering of the NASS states 
(\ref{ascapf}), which vanish when any $k+1$ particles are brought together. 

We refer to a forthcoming paper \cite{ALS} for further analysis of the  
spin-charge separated qH states for general $k$.

\section{Quasi-holes over paired and clustered qH states}
In a BCS superconductor, where electrons are paired up, the 
fundamental flux quantum is reduced to $\frac{1}{2} \Phi_0
= \frac{h}{2e}$. The same phenomenon occurs in the
paired and clustered qH states, and this means that
inserting a single flux quantum $\Phi_0$ creates more than 
a single quasi-hole. For the spin-polarized states of 
\cite{RR1} the number of quasi-holes is given by
$n = k \Delta N_\phi$, where $\Delta N_\phi$ is the
number of excess flux quanta\footnote{Note 
that we adopt a slightly different notation than the
one used in \cite{RR1,GR}. Here, $n$ denotes the number of 
quasi-holes, rather than the number of excess flux quanta.}. 
For the spin-singlet states of \cite{AS1}, this relation becomes
$n^\ua+n^\da = 2k \Delta N_\phi$, were $n^{\ua,\da}$ denotes the 
number of spin-up and down quasi-holes, respectively.

The quasi-holes over the paired and clustered qH states carry 
fractional charge and satisfy non-abelian braid statistics. They 
can be studied with the help of the associated CFT. The wave 
functions of states in the presence of quasi-holes are 
obtained by inserting into the CFT correlators the appropriate 
quasi-hole operators, which consist of a vertex operator part 
and a spin field of the parafermion theory. In the case of the 
MR state, this is the spin field $\sigma$ of the Ising model
and the quasi-hole operator becomes
$\sigma(w):\exp(\frac{i}{2\sqrt{M+1}}\varphi_c):(w)$.

The non-abelian statistics have their origin in the
non-trivial fusion rules of the parafermion spin fields. 
In general, there is more than one way to fuse the fields 
in the correlator to the identity; for $n$ spin fields the 
number of ways will be denoted by $d_n$. The braiding of 
$n$ quasi-holes is then represented by a matrix of size $d_n 
\times d_n$. 

\subsection{Braid statistics} 

The simplest example that exhibits the non-abelian 
braiding is the situation where 4 quasi-holes are added
to the MR state. In this case  $d_4=2$,
so there are two distinct states which, following \cite{NW},
we write as $\Psi^{(4{\rm qh},0)}$ and $\Psi^{(4{\rm qh},\frac{1}{2})}$.
Starting from the state $\Psi^{(4{\rm qh},0)}$, and braiding two
of the particles, we find the following transformation
\be
\Psi^{(4{\rm qh},0)} \rightarrow
\frac{ e^{ \frac{i \pi}{4} }}{\sqrt{2}}
\left( \Psi^{(4{\rm qh},0)} + \Psi^{(4{\rm qh},\frac{1}{2})}
\right) \ .
\end{equation}
Wave functions for the MR state with $n$ quasi-holes 
can be written as \cite{RR1}
\begin{eqnarray} 
\label{pfqh}  
&& \wtPsi_{\rm MR, qh} (z_1,\ldots,z_N;w_1,\ldots,w_n) = 
\frac{1}{2^{(N-F)/2}(N-F)/2!} \prod_{i<j}(z_i-z_j)^{M+1}
\nonu 
&& \times 
\sum_{\sigma \in S_N} {\rm sgn} (\sigma) \prod_{k=1}^F
z_{\sigma(k)}^{m_k} \prod_{l=1}^{(N-F)/2}
\frac{\Phi(z_{\sigma(F+2l-1)},z_{\sigma(F+2l)};w_1,\ldots,w_n)}
{(z_{\sigma(F+2l-1)}-z_{\sigma(F+2l)})} \ , 
\nonu
\end{eqnarray}
where 
\begin{equation}
\Phi (z_1,z_2;w_1,\ldots,w_n)
= \frac{1}{((n/2)!)^2} \sum_{\tau \in S_n} \prod_{r=1}^{n/2}
(z_1-w_{\tau(2r-1)}) (z_2-w_{\tau(2r)}) \ .
\end{equation}
The integers $m_1,\ldots,m_F$ must be chosen such that they satisfy
$0 \leq m_1 < m_2 < \dots < m_F \leq \frac{n}{2}-1$, giving rise to
a degeneracy $ d^{(F)}_n = \left( \begin{array}{c} \frac{n}{2} \\ 
F \end{array} \right)$. The number $F$ is interpreted as the number 
of unpaired electrons in the excited state. 

The braid matrices for $n$ quasi-hole excitations were obtained
by Nayak and Wilczek \cite{NW}, who showed a direct connection with 
the rotation matrices of the group $SO(2n)$. We refer to \cite{SlB}
for more general results on braid matrices.

\subsection{Quasi-hole counting formulas}

The CFT approach to the excited state wave functions and
their braid properties is highly efficient. One would like
however, to `keep both feet on the ground' and understand
the fundamental degeneracies that characterize the non-abelian
statistics in a more direct way. This can be done by selecting 
a (ultra-local) hamiltonian that has the qH state as its 
ground state, and then (numerically) studying the spectrum of 
excited states.

These numerical computations are most easily performed by
studying a small number of particles in a spherical geometry.
By tuning the number of flux quanta to the value
$N_\phi = \frac{1}{\nu} N - S$, where $S$ 
is the so-called shift \cite{We2}, one realizes the qH state 
as the unique ground state. Cranking up the number $N_\phi$ 
and performing a numerical diagonalization, one obtains 
characteristic degeneracies for quasi-hole excitations. 

Following \cite{RR1}, we first explain the counting of 
degeneracies for the case of the MR state.
To understand the degeneracies of quasi-hole excitations
on the sphere, two effects should be taken into account.
The first is a choice of fusion path or, equivalently,
a choice of numbers $F$ and $m_1,\ldots,m_F$ in the formula
(\ref{pfqh}). The second effect is the so-called orbital degeneracy:
the quasi-holes are not localized on the sphere, but can
occupy one of a finite number of available orbitals, each of
which is characterized by a definite angular momentum $L_z$.
These orbital degeneracies are well-known from the
analysis of integer and abelian qH states.

For the MR state, the orbital degeneracy factor depends 
on the number $F$ of unpaired electrons. Fixing this number $F$, 
we have $d_n^{(F)}$ different choices for the quasi-hole 
wave function. To each of those we can associate \cite{RR1} 
an orbital degeneracy factor equal to 
$\left( \begin{array}{c} \frac{N-F}{2}+n \\ n \end{array} \right)$.
Putting it all together, we have the following total degeneracy
\begin{equation}
\#(N,n) = 
\sum_F \left( \begin{array}{c} \frac{n}{2} \\ F \end{array} \right)
\left( \begin{array}{c} \frac{N-F}{2}+n \\ n \end{array} \right) \ ,
\end{equation}
in agreement with numerical results \cite{RR1}. The degeneracies
$d_n$ relevant for a situation where $n$ quasi-holes are at fixed
positions are recovered by suppressing the orbital factors,
$d_n = \sum'_{F} d_n^{(F)} = 2^{n/2-1}$,  where the sum is over 
even (odd) $F$ for $N$ even (odd). This number is in agreement with 
a direct count of the number of fusion paths of $n$ Ising spin 
fields \cite{NW}.

For the more general clustered qH states, the 
degeneracies have basically the same form: an orbital 
part and an intrinsic part, stemming from the non-trivial 
fusion rules. The difference is however, that we can not rely
on explicit wave functions to handle the intrinsic degeneracy.
One can work around this by extracting from the parafermion 
CFT the relevant combinatorial factors, using the methods
put forward in \cite{Sc1,BS}.

For the RR states, the counting has been worked out 
in \cite{GR}, with the result
\begin{equation}
\#(N,n;k) = 
\sum_{F}
\left \{
\begin{array}{c}
n  \\ F \\
\end{array} \right \}_k
\left(
\begin{array}{c}
\frac{N-F}{k} + n \\ n \\
\end{array} \right) \ ,
\end{equation}
with $n$ the number of quasi-holes, $n = k \Delta N_\phi$. 
The symbols $\{^n_F\}_k$ represent the degeneracies
due to the fusion rules. In \cite{BS,GR}, these were described
in terms of recursion relations; explicit formulas (based on
binomials) for general $k$ can be found in \cite{AS2}. 
The sum $d_n = \sum_{F} \{^n_F\}_k$, which equals 
the total number of fusion paths for the spin fields 
contained in $n$ quasi-hole operators, sets the dimension of
the braid matrices for braiding 2 out of the $n$ quasi-particles.

For the NASS states of \cite{AS1}, the
counting goes along the same lines, with the additional
complication that we have to deal with two spin components,
which are combined in a non-trivial way. This is reflected
in the counting formulas by a doubling of the number
of binomial factors. By inserting an amount $\Delta N_\phi$
of extra flux, one creates quasi-holes, which can have
either spin. The total number of quasi-holes is fixed,
$n^\ua+n^\da = 2k\Delta N_\phi$. The symbols $\{ \}_k$ now
depend on four parameters
$\{^{n^\ua \; n^\da}_{F_1 \; F_2} \}_k$ and we have
$d_{n^\ua,n^\da} = \sum_{F_1,F_2}
\{^{n^\ua \; n^\da}_{F_1 \; F_2} \}_k$.
The case $k=2$ has been worked out 
in detail in \cite{ARRS}, where the results are checked against 
numerical data. Explicit results for the symbols $\{ \}_k$ 
can be found in \cite{AS2}. The counting of quasi-hole degeneracies 
over the spin-charge separated spin-singlet states will be discussed
in \cite{ALS}. 

The numbers $d_n$ (for both spin-polarized and spin-singlet 
states) are easily extracted from the known fusion rules of the
$\widehat{su}(2)_k$ and $\widehat{su}(3)_k$ CFTs. For both
$\widehat{su}(2)_3$ and $\widehat{su}(3)_2$ the numbers $d_n$
are Fibonacci numbers. The asymptotic behavior for $n \to \infty$
is found to be
\be
d_n \sim [2 \cos\frac{\pi}{k+2}]^n 
\end{equation}
for the RR states, and
\be
d_p \sim [1+2 \cos\frac{2\pi}{k+3}]^p 
\end{equation}  
for the NASS states, where $p=n^\ua+n^\da$.

\subsection{Exclusion statistics and $K$ matrix structure}

In \cite{ABGS} a proposal was made for a $K$-matrix structure
of the paired and clustered qH states discussed in this paper. 
It was
established that the exclusion statistics of edge excitations
over these states (in a suitable basis) can be captured
by a statistics matrix of the form (\ref{eq:GKK}), 
supplemented by a prescription that some of the particles 
described by this matrix be viewed as pseudo-particles.
We refer to the first paper of \cite{ABGS} for a physical
picture underlying these $K$-matrices, and to the second 
paper of \cite{ABGS} for mathematical details.



\begin{acknowledgments}

K.S. thanks A.~Cappelli and G.~Mussardo for putting together 
a most inspiring Workshop. We acknowledge collaboration with
A.W.W.~Ludwig, N.~Read, E.H.~Rezayi, and stimulating 
discussions with A.~Cappelli and I.T.~Todorov.
This research is supported in part by the Foundation FOM of 
the Netherlands and by the Netherlands Organisation for 
Scientific Research (NWO).

\end{acknowledgments}



\end{document}